\documentclass[acmsmall,screen,nonacm]{acmart}
\usepackage{algorithm}
\usepackage{algpseudocode}
\usepackage{graphicx}
\usepackage{textcomp}
\usepackage{xcolor}
\usepackage{tcolorbox}
\usepackage{hyperref}
\usepackage{wrapfig}
\usepackage{multirow}
\usepackage{makecell}
\usepackage{listings, listings-rust}

\definecolor{prompt_bg}{RGB}{252,255,221}
\definecolor{prompt_title}{RGB}{0,51,102}
\definecolor{Rust}{RGB}{249,104,21}
\definecolor{Go}{RGB}{0,0,255}
\definecolor{offwhite}{RGB}{255,255,242}
\definecolor{GoBackground}{RGB}{200, 230, 240}
\definecolor{RustBackground}{RGB}{255, 235, 210}

\def\BibTeX{{\rm B\kern-.05em{\sc i\kern-.025em b}\kern-.08em
    T\kern-.1667em\lower.7ex\hbox{E}\kern-.125emX}}

\newtheorem{definition}{Definition}

\newcommand{\ourtool}{\textbf{Oxidizer}}

\newcommand{\go}[0]{Go}
\newcommand{\rust}[0]{Rust}
\newcommand{\rustcode}[1]{\textbf{\color{Rust}\small#1}}
\newcommand{\gocode}[1]{\textbf{\color{Go}\small#1}}

\newcommand{\irule}[2]{\mkern-2mu\displaystyle\frac{#1}{\vphantom{,}#2}\mkern-2mu}

\begin{document}

\title{Scalable, Validated Code Translation of Entire Projects using Large Language Models}

\author{Hanliang Zhang}
\affiliation{%
  \institution{University of Bristol}
  \city{Bristol}
  \country{UK}}
\email{hanliang.zhang@bristol.ac.uk}

\author{Cristina David}
\affiliation{%
  \institution{University of Bristol}
  \city{Bristol}
  \country{UK}}
\email{cristina.david@bristol.ac.uk}

\author{Meng Wang}
\affiliation{%
  \institution{University of Bristol}
  \city{Bristol}
  \country{UK}}
\email{meng.wang@bristol.ac.uk}

\author{Brandon Paulsen}
\affiliation{%
  \institution{Amazon, Inc.}
  \city{Virginia}
  \country{US}}
\email{bpaulse@amazon.com}

\author{Daniel Kroening}
\affiliation{%
  \institution{Amazon, Inc.}
  \city{Seattle}
  \country{US}}
\email{dkr@amazon.com}

\begin{abstract}
Large language models (LLMs) show promise in code translation due to their ability to generate idiomatic code.
However, a significant limitation when using LLMs for code translation is scalability: existing works have shown a drop in translation success
rates for code exceeding around 100 lines. We overcome this limitation by developing a modular approach to translation,  where we partition the code into small code fragments which can be translated independently and semantically validated (that is, checking I/O equivalence). When this approach is applied naively, we discover that LLMs are unreliable when translating features of the source language that do not have a direct mapping to the target language, and that the LLM often gets stuck in repair loops when attempting to fix errors. To address these issues, we introduce two key concepts: (1) \emph{feature mapping}, which integrates predefined translation rules with LLM-based translation to guide the LLM in navigating subtle language differences and producing semantically accurate code; and (2) \emph{type-compatibility}, which facilitates localized checks at the function signature level to detect errors early, thereby narrowing the scope of potential repairs.
We apply our approach to translating real-world Go codebases to Rust, demonstrating that we can consistently generate reliable Rust translations for projects up to 6,600 lines of code and 369 functions, with an average of 73\% of functions successfully validated for I/O equivalence, considerably higher than any existing work.


\end{abstract}
\maketitle

% ...





\section{Introduction}
Code translation 
has many important practical applications. For example, a developer may wish to modernize their application~\cite{modernization, developersperceive}, they may wish to maintain a client SDK in multiple languages~\cite{awssdk, googlesdk, azuresdk}, or they may wish to obtain the benefits of another language~\cite{tractor, c2rust}. The Rust language has received special attention as a target language for code translation because it achieves a strong balance between safety and performance~\cite{tractor}. However, manual code translation is tedious and error prone, hence automated code translation would save developer time and energy.

Prior work in automated translation to Rust falls into two categories: rule-based (sometimes called ``symbolic'')~\cite{c2rust, zhang2023ownership} and machine-learning (ML)-based (typically using LLMs)~\cite{tang-etal-2023-explain,RoziereLachaux2020,RoziereZhang2022,szafraniec2022code,yin2024rectifiercodetranslationcorrector}. Rule-based approaches have the advantage of being semantically correct (i.e. I/O equivalent) by construction, but often produce non-idiomatic and unmaintainable code~\cite{PanICSE24, eniser2024translatingrealworldcodellms, yang2024vertverifiedequivalentrust}. Conversely, ML-based approaches are usually more idiomatic and maintainable, but come without correctness guarantees. 

Recent efforts have combined LLM-based translation with a validation step such as running unit tests, differential fuzzing~\cite{eniser2024translatingrealworldcodellms}, or formal verification~\cite{yang2024vertverifiedequivalentrust} to ensure semantic correctness. If validation fails, a new translation is generated. However, these approaches have primarily focused on translating competitive programming-style programs or small code snippets from real-world projects and struggle with translating entire projects. This limitation arises because the context windows of many state-of-the-art LLMs are not large enough to encompass entire projects. Even when the context window suffices, the probability of errors increases exponentially with the amount of code being translated~\cite{eniser2024translatingrealworldcodellms}.

In this work, we aim to move beyond code snippet translation by designing an LLM-based translation approach capable of handling \emph{entire projects} and producing \emph{validated translations}.
A~natural approach to scale to entire projects, taken by us and work parallel to ours~\cite{ibrahimzada2024repositorylevelcompositionalcodetranslation, shiraishi2024contextawarecodesegmentationctorust}, is to partition the project into smaller code fragments (e.g. functions, type definitions) that can be translated individually. We can compute a dependency graph between the fragments, and translate them in post-order, which allows us to incrementally translate and validate the semantic correctness of each translated fragment before translating the next fragment. Moreover, we can provide the LLM with context about translations of previous fragments to aid in translating the current fragment. 

However, this baseline approach alone still struggles to produce semantically correct translations. While the parallel works have achieved high compilation success rates~\cite{ibrahimzada2024repositorylevelcompositionalcodetranslation,shiraishi2024contextawarecodesegmentationctorust}, they report that a large proportion of the original project's test suite, when translated and executed on the translated project, results in runtime errors. This highlights the significant gap between generating compilable code and producing code that actually preserves the original semantics.

Our observation is that any mistake made by the LLM when translating a code fragment can have cascading effects on subsequent translations. First, the LLM often prioritizes generating translations that compile alongside the erroneous fragment, rather than ensuring I/O equivalence with the original code. Second, even when a semantic check identifies the mistake, it may not do so immediately---the error might only be triggered when the code fragment interacts with other parts of the project. The necessary repairs often introduce new errors, affecting other parts of the project, and the LLM can become trapped in a repair loop. This can delay or halt translation progress altogether.

Our solution is to provide specific guidance to the LLM in order to minimize the number of errors the LLM makes from the outset. Next, we outline the strategies we use to achieve this.

\paragraph{\textbf{Predefined Mapping Rules Between Source and Target Language Features}}
We find that LLMs often make errors when (potentially subtle) differences exist between a feature in the source language and the immediately apparent equivalent in the target language. For example, the \go{} and \rust{} languages have different idioms for error handling; \go{} interfaces are structural, while \rust{} traits are nominal; and global variables in \go{} allow dynamic initialization, whereas \rust{} does not. Without guidance on translating these language features, the LLM frequently makes mistakes.

To prevent the LLM from repeating these common errors, we propose using predefined translation rules that map source language features to their counterparts in the target language. As we still want to take advantage of the LLM's ability to generate idiomatic code, we introduce a combined approach that integrates rule-based and LLM-based translation. This method instructs the LLM on which rule to apply and performs static checks on the resulting translation to ensure compliance with these rules.

Throughout the paper, we refer to this technique as \emph{feature mapping}.  When a feature in the source language can be mapped to the target language in multiple ways, feature mapping allows developers to select their preferred mapping and ensures its consistent application throughout the translation process. 

\paragraph{\textbf{Type-Compatibility-Driven Translation}}
We draw inspiration from language interoperability~\cite{multi-language-semantics} and define a set of criteria for \emph{type-compatibility}. This concept ensures that a concrete value of a type in the source language can be converted to a concrete value of the translated type in the target language. Type-compatibility is significant because it is required for I/O equivalence between the source and target code. Moreover, a type-compatibility check is simpler than a I/O equivalence check. It is localized to the signature of the newly translated function, does not require access to the rest of the code, and can be performed immediately after a function's translation. This check can quickly identify mistakes in the function's signature, such as errors in the number or type of arguments. Undetected, these mistakes will cause the unit tests to fail with runtime errors.

Accordingly, we organize the translation process into two phases: a type-driven phase that focuses on producing a type-compatible translation of the project, and a semantics-driven phase that aims to establish I/O equivalence between each original function and its translation. Both type-compatibility and I/O equivalence are evaluated using the source project's test suite.
\\

We implement these strategies in a translation tool called \ourtool{}. While the fundamental ideas proposed in our work are agnostic to the programming language, \ourtool{} specifically targets translating \go{} projects into \rust{} projects. We choose this language pair for several reasons. First, there is a strong push to use memory safe languages like Rust~\cite{tractor}. Second, \go{} and \rust{} operate a similar level of abstraction, and they are often used for similar tasks. Third, \rust{} offers performance and safety benefits over \go{}. Specifically, (safe) \rust{} eliminates data races, and it does not have the overhead of a runtime. Thus there is a strong argument to be made for re-writing \go{} code to \rust{}.

\emph{\textbf{Results.}}
We evaluate \ourtool{} on seven open-source \go{} projects covering diverse use cases, including banking transactions, statistical analysis, and string algorithms. The biggest project has 6.6K lines of code with 369 functions. To our knowledge, this is the largest project translated to Rust using LLMs.

Our results demonstrate that our approach reliably generates translations that pass the \rust{} compiler. Crucially, we successfully validate the I/O equivalence for an average of 73\% of the translated functions (ranging from 63\% to 86\%). This is significantly higher than parallel efforts on full project translation. Shiraishi and Shinagawa~\cite{shiraishi2024contextawarecodesegmentationctorust} focused on generating compilable Rust code and reported that most unit tests crashed on their translations. Ibrahimzada et al.~\cite{ibrahimzada2024repositorylevelcompositionalcodetranslation} validated I/O equivalence for an average of 25.8\% of the translated functions (corresponding to 45.9\% of the functions actually covered by unit tests) and noted that semantic checks resulted in runtime errors for 24.7\% of functions. By contrast, we experienced \emph{no runtime errors}---every failing unit test failed due to assertion errors when using our approach.
We also show that both feature mapping and type-compatibility are critical for reliably producing translations of entire projects. 

These findings are highly encouraging, especially considering that parallel research~\cite{ibrahimzada2024repositorylevelcompositionalcodetranslation}, which incorporated a user study, found that even partial translations with considerably lower I/O equivalence rates (25.8\% overall and 45.9\% for covered functions) can substantially reduce the time developers spend translating a codebase.


\emph{\textbf{Contributions.}}

$\bullet$ We present an LLM-based translation approach for entire projects, achieving a high 73\% rate of functions validated for I/O equivalence.

$\bullet$ We introduce feature mapping, a technique that combines predefined translation rules with LLM-based translation to guide the LLM in handling subtle language differences, ensuring semantically correct translations.

$\bullet$ We introduce type-compatibility as a prerequisite for I/O equivalence, enabling localized checks at the function signature level to identify errors early and prevent runtime failures.

$\bullet$ We implement our proposed approach in a tool \ourtool{}, and demonstrate its ability to reliably produce useful \rust{} translations on seven open-source \go{} projects.

\section{Overview of \ourtool{}} \label{sec:overview}
Throughout this paper we use \gocode{blue} to denote code in the source language (Go in our case), and \rustcode{orange} to denote code in the target language (Rust in our case).
We illustrate our approach using the example in Figure~\ref{fig:go_overview}, which is a contrived version of code extracted from one of our real-world benchmarks for validating banking transactions~\cite{ach}. The example consists of three files: ``types.go'', which declares the type \gocode{EntryDetail}; ``globals.go'', which defines the global variable \gocode{specialNumber} initialized by a call to \gocode{setupSpecialNumber}; and ``validator.go'', which contains the definition of the function \gocode{Validate}.
We assume all three files are located in the same package.
\begin{figure}[!h]
    \centering
    \begin{minipage}{.47\textwidth}
        \begin{lstlisting}[language=go, basicstyle=\tiny\ttfamily,escapechar=\&,backgroundcolor=\color{GoBackground}]
// in file globals.go
var specialNumber int = setupSpecialNumber()

func setupSpecialNumber() int { ... }

// in file types.go
type EntryDetail struct {
	Addenda05 []int
}

&\newline&
&\newline&
&\newline&
&\newline&
&\newline&
\end{lstlisting}
        \end{minipage}
    \begin{minipage}{0.5\textwidth}
        \begin{lstlisting}[language=go, basicstyle=\tiny\ttfamily,escapechar=\&,backgroundcolor=\color{GoBackground}]
// in file validator.go        
func Validate(entry *EntryDetail, length int) bool {
    if entry.Addenda05 != nil {
        // length check
        if len(entry.Addenda05) != length {
            return false
        }
        // search for special number
        for _, r := range entry.Addenda05 {
            if r == specialNumber { return true; }
        }
        return false;
    }
    return false;
}
\end{lstlisting}
\end{minipage}%
\caption{Source Go code consisting of three files: globals.go, types.go, and validator.go}
    \label{fig:go_overview}
\end{figure}

Translation happens in three phases: \textbf{project partitioning}, \textbf{type-driven translation}, and \textbf{semantics-driven translation}. 

\subsection{Project partitioning}
Our approach first partitions the project into code fragments to be translated individually. We create a fragment for each function/method, type definition (i.e. struct or interface), and global variable declaration. Specifically, we create fragments for: 
the type definition \gocode{EntryDetail}, the global variable \gocode{specialNumber}, the function \gocode{setupSpecialNumber}, and the function \gocode{Validate}. 
These fragments are then organized into a \emph{dependency graph}, where the graph edges capture a code fragment's usage of other type definitions, global variables, and functions/methods. The relations we have are: 
\gocode{specialNumber} depends on \gocode{setupSpecialNumber};
\gocode{Validate} depends on \gocode{EntryDetail} and \gocode{specialNumber}.

\subsection{Type-Driven Translation}
Next, we move into the type-driven translation phase. We iterate over the code fragments in post-order according to the dependency graph, and prompt the LLM for a translation of each code fragment.
The goal of our type-driven translation phase is to get a compiling translation, and to catch certain errors related to types.
We defer checking I/O equivalence until the next phase.

Assume we have a translation for the function \gocode{setupSpecialNumber}, and are now translating the global variable declaration \gocode{specialNumber}. A translation obtained from Anthropic's Claude~3 Sonnet is given in
\autoref{fig:rust_overview_wrong}, where the global variable definition is translated to a global \rustcode{static} definition.
\begin{figure}[!h]
    \centering
    \begin{minipage}{.47\textwidth}
        \begin{lstlisting}[language=rust, basicstyle=\tiny\ttfamily,escapechar=\%]
// in file globals.rs
static special_number: i32 = setup_special_number()

fn setup_special_number() -> i32 { ... }

// in file types.rs
struct EntryDetail {
    addenda05: Vec<i32>,
}

%\newline%
%\newline%
%\newline%
%\newline%
%\newline%
\end{lstlisting}
        \end{minipage}
    \begin{minipage}{0.5\textwidth}
        \begin{lstlisting}[language=rust, basicstyle=\tiny\ttfamily,escapechar=\%]
// in file validator.rs        
fn validate(length: u32, entry: &EntryDetail)->bool {
    if entry.addenda05.iter().all(|x| *x!=0) {
        // length check
        if entry.addenda05.len() != length as usize {
            return false
        }
        // search for special number
        for r in entry.addenda05.iter() {
            if *r == special_number { return true }
        }
        return false
    }
    return false
}
\end{lstlisting}
    \end{minipage}%
    \caption{Incorrect Rust translation for the Go code in \autoref{fig:go_overview}}
        \label{fig:rust_overview_wrong}
    \end{figure}
\paragraph{\textbf{Challenge: LLM Incorrectly Maps Source Language Features to Target Language Features}}
While this may seem like an intuitive translation, it will trigger a compilation error because, while  
a global variable can be initialized by arbitrary functions in Go, \rustcode{static} global variables in Rust can only be initialized by functions that can be evaluated at compile-time. 

The difficulty in correctly mapping Go global variable definitions to Rust highlights a broader challenge: LLMs often struggle with accurately translating language features, particularly when there are subtle syntactic or semantic differences between constructs in the source and target languages~\cite{yin2024rectifiercodetranslationcorrector}. In such cases, LLMs may incorrectly mimic source language syntax in the target language or misalign source APIs with target ones, resulting in code that fails to compile or is semantically not equivalent.

\paragraph{\textbf{Solution: Feature Mapping Rules}}
To address this issue, we propose to use predefined feature mapping rules to constrain the LLM-based translation. These rules are tailored to a specific pair of source and target languages. 
For the particular case of global variable definition, 
several workarounds are available, utilizing constructs like \rustcode{Lazy} and \rustcode{lazy\_statics}, which defer function evaluation until its first access.
We pick the former option, and instruct the LLM on the desired target feature. Consequently, Claude 3 generates the correct translation for \gocode{special\_number} shown in \autoref{fig:rust_overview_correct}. To ensure rules are followed, we pair each translation rule with a set of static, symbolic checks, and re-query the LLM for a new translation if these checks fail.
\begin{figure}[!h]
    \centering
    \begin{minipage}{.47\textwidth}
        \begin{lstlisting}[language=rust, basicstyle=\tiny\ttfamily,escapechar=\%]
// in file globals.rs
static special_number: Lazy<i32> =
        Lazy::new(|| setup_special_number());
fn setup_special_number() -> i32 { ... }
// in file types.rs
pub struct EntryDetail {
    addenda05: Option<Vec<i32>>,
}
\end{lstlisting}
        \end{minipage}
    \begin{minipage}{0.5\textwidth}
        \begin{lstlisting}[language=rust, basicstyle=\tiny\ttfamily,escapechar=\%]
// in file validator.rs        
use crate::types::EntryDetail;
use crate::globals::special_number;
fn validate(entry: &EntryDetail, length: u32)->bool {
    if entry.addenda05.is_some() {
        // same as %\autoref{fig:rust_overview_wrong}%
        ...
%\newline%
\end{lstlisting}
    \end{minipage}%
    \caption{Correct Rust translation for the Go code in \autoref{fig:go_overview}}
        \label{fig:rust_overview_correct}
    \end{figure}

\paragraph{\textbf{Challenge: Incompatible Types}}
Now assume we move on to translating the \gocode{EntryDetail} type definition, and obtain the translation given in \autoref{fig:rust_overview_wrong}. While the translation compiles successfully, errors remain. In the original Go code, the \gocode{Addenda05} field of \gocode{EntryDetail} has type \gocode{[]int}, which is nullable, whereas in the Rust translation, the corresponding field \rustcode{addenda05} has type \rustcode{Vec<i64>}, which is not nullable.
This translation is semantically inequivalent to the original, and will cause semantic validation to fail in the next phase of translation. Moreover, if we continue translating the rest of the code fragments, they will use this incorrectly translated type definition, which will increase the number of repairs we will need to make in the future. This is illustrated by the translation for the function \rustcode{validate} in \autoref{fig:rust_overview_wrong}, which checks that all values in \rustcode{entry.addenda05} are non-zero, whereas in the original Go function, \gocode{Validate} checks \gocode{entry.Addenda05 != nil}.


\paragraph{\textbf{Solution: Type-Compatibility Checks}}
To catch these semantic in-equivalences early on, we define a notion of \textit{type-compatibility}. 
Type-compatibility ensures that a concrete value of a type in the source language can be converted to a concrete value of the corresponding type in the target language, a property necessary for validating I/O equivalence between functions in the source language and target language. As detailed later in the paper, we verify this using execution snapshots (i.e. I/O pairs) for each function, which were extracted from the project's test suite.

The translation given in \autoref{fig:rust_overview_wrong} fails to satisfy this definition because \gocode{null} can inhabit the \gocode{Addenda05} field of \gocode{EntryDetail} in Go source, but the same is not true for the Rust translation. Although this error would eventually be detected by the semantic validation check in the next phase, this is delayed until \gocode{Validate} is translated (as semantic validation checks that functions in the original and translated code are I/O equivalent). By that stage, additional code has been generated, making the error more difficult to diagnose: it could mistakenly appear as an issue with the \rustcode{validate} translation rather than with \rustcode{EntryDetail}. Also, fixing the error would require backtracking. With type-compatibility checks, we can catch this error immediately after generating \rustcode{EntryDetail}, avoiding these complications.

Following the failure of the type compatibility check, we re-query the LLM, and obtain the correct translation \rustcode{EntryDetail}
in \autoref{fig:rust_overview_correct}, which wraps \rustcode{Option} around \rustcode{Vec}. 
We also perform type compatibility checks for function/method signatures, which assert the compatibility of their constituent types. For our example, the signature of \gocode{Validate} would also fail the type compatibility check with its Rust translation \rustcode{validate} because the two arguments are swapped. After re-querying, Claude 3 returns the correct translation in \autoref{fig:rust_overview_correct}. 

\paragraph{\textbf{Visibility modifiers and import statements.}} LLMs often struggle to generate accurate visibility modifiers and import statements when the organizational principles of the source and target languages differ. To address this, we use a dependency analysis to post-process all translations, ensuring the correct visibility modifiers and import statements are applied, as demonstrated in~\autoref{fig:rust_overview_correct}.


\subsection{Semantics-Driven Translation}
After the type-driven translation phase, we assume we have a \emph{type-compatible project translation}, which compiles successfully, and all type definitions and function/method signatures in the source project are type-compatible with their translations, as shown in \autoref{fig:rust_overview_correct}. 
We then move to the semantics-driven translation phase, during which we test the translated functions for I/O equivalence with the corresponding functions in the source project. 
We check I/O equivalence of each translated global definition and each function based on input-output examples extracted by running the source project's unit tests. 
To isolate a particular function translation during I/O equivalence validation, we mock the translated function's callees, which ensures that I/O equivalence failures are due to bugs in the function under test, and allows us to focus our repair efforts on that function.
In this example, the translation successfully passes the semantic validation checks.

\section{Project Partitioning}
The first step in our translation procedure is partitioning the source project into fragments that can be translated individually.
We describe our partitioning strategy for Go, though the same approach will apply for most source languages.
For simplicity, we assume the project follows the
simplified FeatherweightGo~\cite{griesemer2020featherweight} specification, as given in~\autoref{golang}. A Global declaration defines a package-level global variable $x$ with type $T$, which is initialized by either a value $v$ or through a function call $f(\overline{v})$. A Type declaration defines a named type $T$ that is either  a struct or an interface, with the latter being a set of function signatures.
A Function/Method declaration defines a function/method named $f$ with its signature and body. In this simplified language specification, we ignore function bodies as they are not important for the purpose of code partitioning (as
we never partition a function's body). A Signature allows multiple input and output types. Notably, the idiomatic Go error handling mechanism makes use of multiple return types (where some correspond to the possible errors). Then, a Go project consists of a set of top-level declarations \gocode{$\overline{D}$} for global variables, types, functions and methods, which can be spread across multiple files. 
We partition the Go project according to these declarations. 
We made this decision as each declaration carries distinct type and semantic information, making it suitable for independent translation and correctness checks, as discussed further in the paper.


\begin{figure}[!htb]
\begin{minipage}{.47\textwidth}
\centering
\small 
\[
\begin{array}{llcl}
\text{Function Name} & & & f, g\\
\text{Field Name} & & & F, G\\
\text{Variable Name} & & & x, y\\
\text{Type Name} & & & T, U, I\\
\text{Value} & & & v, w\\
\text{Global} & & ::= &  \gocode{var}\ x\ T\ =\ v\ |\ f(\overline{v})\\
\text{Type} & & ::= & \gocode{type}\ T\ \ell\\
\text{TypeLiteral} & \ell & ::= & \gocode{struct}\ \{ \overline{F\ T} \}\ |\ \\
&&&\gocode{interface}\ \{ \overline{f\ S} \}\\
\text{Fn} & & ::= & \gocode{func}\ f\ S\ \{\dots\}\\
\text{Method} & & ::= & \gocode{func}\ (x\ T)\ f\ S\{\dots\}\\
\text{Signature} & S & ::= & (\overline{x\ T})\ \overline{U}\\
\text{Declaration} & D & ::= & \text{Global} \ |\  \text{Type}\ |\ \text{Fn}\ |\ \\
&&&\text{Method}\\
\text{Project} & P & ::= & \overline{D}
\end{array}
\]
\caption{Simplified Go specification\label{fig:go-spec}}
\label{golang}
\end{minipage}\begin{minipage}{.57\textwidth}
\centering
\small 
\[
\begin{array}{llcl}
\text{Function Name} & & & f, g\\
\text{Field Name} & & & F, G\\
\text{Variable Name} & & & x, y\\
\text{Type Name} & & & T, U\\
\text{Trait Name} & & & I\\
\text{Value} & & & v, w\\
\text{Static} & & ::= &  \rustcode{static}\ x: T\ =\ v\\
\text{Type} & & ::= & \rustcode{struct}\ T\ \{ \overline{F: T} \}\\
\text{Trait} & & ::= & \rustcode{trait}\ I\ \{ \overline{\rustcode{fn}\ f\ S_m} \}\\
\text{Fn} & & ::= & \rustcode{fn}\ f\ S_f\ \{\dots\}\\
\text{Impl} & & ::= & \rustcode{impl}\ T\ \{ \overline{\rustcode{fn}\ f\ S_m\  \{\dots\}} \}\\
\text{FnSignature} & S_f & ::= & (\overline{x: T}) \rightarrow \overline{U}\\
\text{MethodSignature} & S_m & ::= & (\rustcode{self}, \overline{x: T}) \rightarrow \overline{U}\\
\text{SelfType} & \rustcode{self} & ::= & \rustcode{\&self}\ |\ \rustcode{\&mut self}\ |\ \dots \\
\end{array}
\]
\caption{Simplified Rust specification}
\label{rustlang}
\end{minipage}
\end{figure}

\section{Translating an Individual Code Fragment}
In this section, we describe the translation of an individual code fragment. In the following sections, we explain translating the entire project.

The most critical aspect when translating a code fragment is our \textit{feature mapping rules}. 
As illustrated previously in \autoref{sec:overview}, LLMs don't always generate the semantically correct feature mapping. Our observation is that, while getting these translations correct is challenging for an LLM, a rule-based approach can predefine mappings between features of the source and target languages. 
While we could write purely symbolic translation rules, this approach has two major disadvantages. First, as prior work~\cite{eniser2024translatingrealworldcodellms,PanICSE24, yang2024vertverifiedequivalentrust} has shown, symbolic translation rules often produce unidiomatic code. 
Second, a purely symbolic approach demands an exhaustive set of translation rules covering every feature. This would incur a significant amount of developer effort.


This motivates us to take a hybrid approach that combines an LLM and static analysis to implement feature mapping rules. We express a feature mapping rule as a judgment, and implement it in three parts: (1) a syntactic pattern that detects when the rule applies (i.e. the premise), (2) a natural language description of the translation rule, which is provided to an LLM, and (3) a set of static checks that validate whether the rule was applied correctly (i.e. the conclusion). We explain our feature mapping rules for Go to Rust in \autoref{sec:feature-mapping}. 

\autoref{alg:feature-mapping} describes how we apply our feature mapping rules (and how we obtain candidate translations as well). The algorithm takes a source code fragment to translate \gocode{D}, and a re-query budget $\mathit{requery\_budget}$. It returns a translation of \gocode{D} that satisfies the conclusions of any applicable feature mapping rules.

We first statically analyze \gocode{D} to determine which, if any, feature mapping rules apply, and retrieve the natural language descriptions of the rules, and their conclusions (i.e. validation checks). Next, we compute a summary of the translations for any dependencies of \gocode{D}. This summary is a condensed version of the translated code fragments. It contains function/method signatures without bodies, and it contains type definitions and global variable declarations. We then use both of these to instantiate the LLM prompt, which is given in \autoref{fig:feature-request-prompt}. We then iteratively query the LLM with this prompt to obtain a translation, and check the translation satisfies the feature mapping rules' conclusions. Once the checks are satisfied, we may apply some ad hoc post processing, and we compute the necessary visibility modifiers (i.e. target language-specific import statements). If the $\mathit{requery\_budget}$ runs out, we abort translation, but this did not happen in our experiments.

\begin{algorithm}
\caption{Translation of an individual fragment with feature mapping}\label{alg:features-mapping}
\begin{algorithmic}
  \footnotesize
\Require Source Code Fragment $\gocode{D}$, Requery Budget $\mathit{requery\_budget}$
\Ensure $\rustcode{target\_code}$, Target Language Code that Satisfies Feature Mapping Rules
\State $\mathit{rule\_descriptions}, \mathit{conclusions} \gets \textbf{GetFeatureMappingRules}(\gocode{D})$
\State $\mathit{dependencies} \gets \textbf{GetDependenciesSummary}(\gocode{D})$
\State $\mathit{prompt} \gets \textbf{GeneratePrompt}(\gocode{D}, \mathit{rule\_descriptions}, \mathit{dependencies})$
\State $\mathit{correct\_mapping} \gets \mathit{false} $
\While{$\neg \mathit{correct\_mapping}$}
\State $\rustcode{code} \gets \textbf{QueryLLM}(prompt)$
\State $\textit{correct\_mapping} \gets \textbf{CheckFeatureMapping}(\rustcode{code}, \mathit{conclusions})$
\State $\mathit{requery\_budget} \gets \mathit{requery\_budget} - 1$
\If{$\mathit{requery\_budget} \leq 0 $}
\State \textbf{AbortTranslation()}
\EndIf
\EndWhile
\State $\rustcode{target\_code} \gets \textbf{PostProcessing}(\rustcode{code})$
\State $\rustcode{target\_code} \gets \textbf{SetVisibilityModifiers}(\rustcode{target\_code})$
\State return \rustcode{target\_code}
\end{algorithmic}
\label{alg:feature-mapping}
\end{algorithm}

\begin{figure}
    \centering
    \begin{tcolorbox}[
        colback=prompt_bg,
        colframe=prompt_title,
        subtitle style={boxrule=0.4pt, colback=yellow!50!blue!25!white, colupper=black},
    ]
    \scriptsize
    Below, you are given a fragment of Go code. Your job is to translate it to Rust.\\
    $\$\{\mathit{SOURCE\_CODE\_FRAGMENT}\}$\\
    \\
    \texttt{The dependencies of the above Go code have already been translated to Rust. 
    A condensed version of their translations is given below. 
    Make sure to use them in your translation.}\\
    $\$\{\mathit{TRANLSATED\_DEPENDENCIES}\}$\\
    \\
    \texttt{When translating the fragment of Go code, apply the following translation rules.}\\
    $\$\{\mathit{FEATURE\_MAPPING\_RULES}\}$
    
    \end{tcolorbox}
    \caption{LLM prompt template}
    \label{fig:feature-request-prompt}
\end{figure}

\subsection{Mapping Language Features} \label{sec:feature-mapping}
In the rest of the section, we discuss the Go-to-Rust feature mapping rules for basic features (e.g. variable definitions, struct definitions), interface declarations and error handling.

\subsubsection{Mapping basic features}
\autoref{fig:basic-mapping} illustrates the rules for mapping basic Go features. Each rule's premise states that, given a Go fragment \gocode{D} containing a feature of interest, we query the LLM to produce \rustcode{code}, denoted by $\rightsquigarrow_{\text{LLM}}$. The augmentation of the prompt with the specified feature mapping is implied within $\rightsquigarrow_{\text{LLM}}$. The symbol $\Downarrow$ in the conclusion denotes the check confirming that \rustcode{code} includes the expected Rust feature. If this condition is met, then the fragment \gocode{D} is successfully translated to \rustcode{code}.

For example, the rule (\text{Map-Var-Init}) says that given a Go global variable definition $\gocode{var}\ x = f(\overline{v})$, where the variable is initialized by a function call, we construct a prompt to query the LLM for a translation \rustcode{code}. 
The expected translation for the global variable in \rustcode{code} is a Rust global static definition with its initializer wrapped in \rustcode{Lazy::new}, as checked by $\Downarrow$.
If the check succeeds, then \rustcode{code} is treated as the translation of \gocode{D}. Code wrapped by \rustcode{Lazy::new} is evaluated upon first access, which we consider to be an emulation of Go global \gocode{var} initialization. 
While there are alternative emulation methods, we choose \rustcode{Lazy::new} for its simplicity. However, different developers may customize the rule differently, opting for a different alternative.

Beyond the \rustcode{Lazy::new} wrapper, we don't impose any additional constraints. The LLM is free to generate names (e.g., corresponding to $y$ in the conclusion of the rule), types (e.g., corresponding to~$U$), and initializers (the argument of \rustcode{Lazy::new} is not required to be a function call). 
For idiomaticity purposes, we do not constrain the variable, type and field names: Go prefers Camel Case, whereas Rust prefers Snake Case.

\begin{figure}[h]
\centering
\small 
\[
\begin{array}{cc}
\irule{
\gocode{D}: \gocode{var}\ x\ T = f(\overline{v}) \rightsquigarrow_{\text{LLM}}  \rustcode{code}
}{
\rustcode{code} \Downarrow \rustcode{static}\ y:\rustcode{Lazy<}U\rustcode{>} = \rustcode{Lazy::new(||}\ \dots \rustcode{)} \vdash \gocode{D} \mapsto \rustcode{code}
} &  
\irule{
\gocode{D}: \gocode{type}\ T\ \gocode{struct}\ \{\dots\}
\rightsquigarrow_{\text{LLM}} \rustcode{code}
}{
\rustcode{code} \Downarrow \rustcode{struct}\ U \{\dots\}\vdash \gocode{D} \mapsto \rustcode{code}} \\ \\
(\text{Map-Var-Init}) 
 & (\text{Map-Struct}) \\ \\ 
\irule{
\gocode{D}: \gocode{func}\ (x\ T)\ f\ S\rightsquigarrow_{\text{LLM}}  \rustcode{code}
}{
\rustcode{code} \Downarrow \rustcode{impl}\ Tr\ \{ \rustcode{fn}\ g\ S_m \{\dots\} \} \vdash \gocode{D} \mapsto \rustcode{code}
} &
\irule{
\gocode{D}: \gocode{func}\ f\ S
\rightsquigarrow_{\text{LLM}}  \rustcode{code}
}{
\rustcode{code} \Downarrow \rustcode{fn}\ g\ S_f
 \vdash \gocode{D} \mapsto \rustcode{code}
}  \\ \\
(\text{Map-Method}) 
&
(\text{Map-Fn}) 
\end{array}
\]
\caption{Basic feature mapping}
\label{fig:basic-mapping}
\end{figure}

The remaining rules in Figure~\ref{fig:basic-mapping} outline feature mappings for struct type definitions, functions, and methods: a Go struct type is mapped to a Rust struct type, Go functions translate to Rust free-standing functions, and Go methods map to Rust inherent implementations.
Similar to the (\text{Map-Var-Init}) rule, names and types in the conclusion are left unconstrained.

\subsubsection{Mapping error handling}
\label{sec:error-handling}
Go and Rust adopt drastically different error handling styles. In Go, error handling often involves returning an \gocode{error} as an additional return value, which is then checked by the caller using simple if statements, as illustrated in~\autoref{fig:go_error_handling}.
\begin{figure}[!htb]
    \centering
    \begin{minipage}{.5\textwidth}
        \begin{lstlisting}[language=go, basicstyle=\tiny\ttfamily,escapechar=\&,backgroundcolor=\color{GoBackground}]
func f() (uint32, error) { ... }
func g() error {
    if x, err := f(); err != nil {
        if _, ok := err.(*BatchError); ok {
	       return err
        }
        return BatchError{err}
    }
    ...
}
&\newline&
\end{lstlisting}
        \caption{Source Go snippet}
        \label{fig:go_error_handling}
    \end{minipage}%
    \begin{minipage}{0.5\textwidth}
        \begin{lstlisting}[language=rust, basicstyle=\tiny\ttfamily]
fn f() -> Result<u32, FError> { ... }
fn g() -> Result<(), BatchError> {
    let x = f().map_err(|err| {
        if let Ok(err) = err.downcast::<BatchError>() 
        {
            return err
        }
        return BatchError(err)
    })?;
    ...
}
\end{lstlisting}
        \caption{Incorrect translation to Rust}
        \label{fig:rust_error_handling}
    \end{minipage}
\end{figure}
The builtin \gocode{error} type is an \gocode{interface} type containing exactly one method: \gocode{Error() string}. Although custom error types can be constructed separately, they often implement this interface and are passed around as \gocode{error} rather than concrete types. Then, type assertions are used to recover the concrete underlying type. In this example, the caller \gocode{addendaFieldInclusion} checks whether the error returned by \gocode{checkAddenda02} is an instance of the custom error \gocode{BatchError} (i.e.~ \gocode{err.(*BatchError)}).
By contrast, idiomatic Rust represents potential errors using the \rustcode{Result} type, enabling monadic error propagation using the \rustcode{?}~operator. Rust does not have a built-in error type; instead, idiomatic Rust often involves passing concrete error types, with additional code needed to convert between them as necessary.

\autoref{fig:rust_error_handling} presents a translation generated by LLMs for the code in Figure~\ref{fig:go_error_handling}. Although this code looks idiomatic, it does not compile due to the erroneous emulation of the source syntax: the Go type assertion \gocode{err.(*BatchError)} is translated to  \rustcode{err.downcast::<BatchError>()}, which is a compilation error since \rustcode{FError} does not have a \rustcode{downcast} method.

There are various ways to map error handling between Go and Rust, giving developers flexibility to choose their preferred approach. In this work, we chose to predefine a unified concrete Rust error type.
We selected \rustcode{anyhow::Error} from a widely used third-party Rust library for idiomatic error handling~\cite{anyhow}.
 The advantage of \rustcode{anyhow::Error} is that it acts as a lightweight wrapper for any type implementing the built-in \rustcode{std::error::Error} trait, facilitating easy conversion between such types and enabling instance checking for custom error types that also implement \rustcode{std::error::Error}. Additionally, the trait \rustcode{std::error::Error} relies on \rustcode{std::fmt::Display} for pretty-printing of error messages, which provides a natural correspondence for the \gocode{Error() string} method.

We define the feature mapping rules for error handling in~\autoref{fig:error-handling-mapping}. The rule (Map-Custom-Error) instructs the LLM to translate the \gocode{error} interface implementation into three Rust trait implementations so that the translated custom error implements \rustcode{std::error::Error}. The rule (Map-Error-Handling-Fn) ensures that  error handling Go functions are translated to idiomatic Rust, where the error type is constrained to be the predefined \rustcode{anyhow::Error}.

\begin{figure}[h]
\centering
\small 
\[
\begin{array}{cc}
\irule{
\gocode{D}: \gocode{func}\ (t\ T)\ \gocode{Error}()\ \gocode{string}\ \{\dots\}
\rightsquigarrow_{\text{LLM}} \rustcode{code}
}{
\rustcode{code} \Downarrow 
\begin{array}{c}
    \rustcode{impl}\ \rustcode{Debug}\ \rustcode{for}\ U\ \{\dots\}\\
    \rustcode{impl}\ \rustcode{Display}\ \rustcode{for}\ U\ \{\dots\}\\
    \rustcode{impl}\ \rustcode{std::error::Error}\ \rustcode{for}\ U\ \{\dots\}
\end{array}
\vdash \gocode{D} \mapsto \rustcode{code}
} & (\text{Map-Custom-Error})\\ \\ 
\irule{
\gocode{D}: \gocode{func}\ f\ (\overline{x\ T})\ (\overline{U}, \gocode{error})
\rightsquigarrow_{\text{LLM}}  \rustcode{code}
}{
\rustcode{code} \Downarrow \rustcode{fn}\ g\ (\overline{x: Tr}) \rightarrow \rustcode{Result<}\overline{Ur}, \rustcode{anyhow::Error}\rustcode{>}
 \vdash \gocode{D} \mapsto \rustcode{code}
} & (\text{Map-Error-Handling-Fn}) \\ \\ 
\end{array}
\]
\caption{Mapping error handling}
\label{fig:error-handling-mapping}
\end{figure}

\subsubsection{Mapping Go interfaces to Rust traits}
\label{sec:interface-mapping}

Go \gocode{interface} and Rust \rustcode{trait} are similar abstractions in the sense that both achieve polymorphism by defining a collection of methods that 
get implemented for different types. \autoref{fig:go_snippet_interface} presents two Go interfaces, \gocode{Batcher} and \gocode{canValidate}, extracted from the realworld benchmark \textbf{ach}.
\begin{figure}[!htb]
    \centering
    \begin{minipage}{.5\textwidth}
        \begin{lstlisting}[language=go, basicstyle=\tiny\ttfamily,escapechar=\&,backgroundcolor=\color{GoBackground}]
type Batcher interface {
    SetHeader(*BatchHeader)
    Validate() error
}
&\newline&
&\newline&
type canValidate interface {
    Validate() error
}
&\newline&
&\newline&
func (t *T) SetHeader(*BatchHeader) { ... }
func (t *T) Validate() error        { ... }
&\newline&
&\newline&
\end{lstlisting}
        \caption{Source Go snippet}
        \label{fig:go_snippet_interface}
    \end{minipage}%
    \begin{minipage}{0.5\textwidth}
        \begin{lstlisting}[language=rust, basicstyle=\tiny\ttfamily]
trait Batcher {
    fn set_header(&mut self, _: Option<BatchHeader>);
    fn validate(&self) -> Result<()>;
}
trait canValidate {
    fn validate(&self) -> Result<()>;
}
impl Batcher for T {
    fn set_header(&mut self, _: Option<BatchHeader>) 
    { ... }
    fn validate(&self) -> Result<()> { ... }
}
impl canValidate for T {
    fn validate(&self) -> Result<()> { ... }
}
\end{lstlisting}
        \caption{Incorrect translation to Rust}
        \label{fig:rust_snippet_interface}
    \end{minipage}
\end{figure}
Given that a Go interface is \emph{structural} and \gocode{Batcher} and \gocode{canValidate} share the method signature for \gocode{Validate}, 
\gocode{Batcher} and \gocode{canValidate} exhibit a sub-interfacing relation. As a result, a value of type \gocode{Batcher} is \emph{assignable} to a variable of type \gocode{canCandidate}. 
Furthermore, the concrete implementation of the \gocode{Validate} method only needs to be provided once.

Figure~\ref{fig:rust_snippet_interface} presents a translation obtained from the LLM for Figure~\ref{fig:go_snippet_interface} using Rust traits.
Conversely to Go interfaces, a Rust \rustcode{trait} is \emph{nominal}, and thus the sub-interfacing relation does not hold. This is a translation error as, for any Go code where a value of type \gocode{Batcher} is assigned to/upcast to a variable of type \gocode{canValidate},
there will not exist any compilable Rust translation. 
Moreover, this translation also poses a challenge for our design based on modular translation and validation of individual functions: all the method implementations for an interface must be translated at the same time in order for the code to pass the compiler check.

To address these challenges, we propose a target feature
that decomposes a trait definition into multiple sub-traits (note that our notion of sub-trait doesn't induce a subtyping relation), each containing a single method signature. Methods that appear in multiple interfaces only need one such sub-trait. Then the main trait (corresponding to the original Go interface) is bounded by all its sub-traits. Figure~\ref{fig:rust_interface_feature} illustrates this solution 
for the traits in Figure~\ref{fig:go_snippet_interface}. 
If we take the example of \gocode{canValidate}, it gets translated to the following  Rust components:
\begin{enumerate}
    \item \emph{main trait} \rustcode{canValidate}, which is bounded by its corresponding sub-trait \rustcode{canValidate\_Validate};
    \item \emph{sub-trait} \rustcode{canValidate\_Validate}, corresponding to method \rustcode{Validate};
    \item implementation of sub-strait \rustcode{canValidate\_Validate} for \texttt{T};
    \item implementation of the main trait for \texttt{T}; this is automatically generated by \ourtool{} as part of \textbf{PostProcessing} in Algorithm~\ref{alg:feature-mapping} once all sub-traits (in this case just \rustcode{canValidate\_Validate}) are implemented.
\end{enumerate}
\begin{figure}[!htb]
    \centering
    \begin{minipage}{.5\textwidth}
        \begin{lstlisting}[language=rust, basicstyle=\tiny\ttfamily]
// main trait
trait canValidate: canValidate_Validate {}
// sub-trait
trait canValidate_Validate {
    fn validate(&self) -> Result<()>;
}
// auto impl
impl<T> canValidate for T
where
    T: canValidate_Validate {}

// main trait
trait Batcher: canValidate_Validate + Batcher_SetHeader + canValidate {}
// sub-trait
trait Batcher_SetHeader {
    fn set_header(&mut self, _: Option<BatchHeader>);
}
\end{lstlisting}
   \end{minipage}%
    \begin{minipage}{0.47\textwidth}
        \begin{lstlisting}[language=rust, basicstyle=\tiny\ttfamily, escapechar=\%]
// auto impl
impl<T> Batcher for T
where
    T: canValidate_Validate + Batcher_SetHeader {}

impl Batcher_SetHeader for T {
    fn set_header(&mut self, _: Option<BatchHeader>) 
    { ... }
}

impl canValidate_Validate for T {
    fn validate(&self) -> Result<()> { ... }
}
%\newline%
%\newline%
%\newline%
%\newline%
\end{lstlisting}
    \end{minipage}
\caption{Correct translation
        of Go \gocode{interface} to Rust \rustcode{trait}}
        \label{fig:rust_interface_feature}
\end{figure}

We follow a similar approach for \gocode{Batcher}, where we reuse the already generated sub-trait \rustcode{canValidate\_Validate} corresponding to method \rustcode{Validate}. The main trait \rustcode{Batcher} is bounded by the sub-traits \rustcode{canValidate\_Validate} and \rustcode{Batcher\_SetHeader}. Additionally, we add an extra bound \rustcode{canValidate} for \rustcode{Batcher} to properly reflect the sub-interfacing relation present in the source Go code. Due to space limitations, we omit the formal rule for interface mapping.

\section{Type-Driven Translation}\label{sec:type-driven}
So far, we have focused on translating individual code fragments. In this section, we shift our attention to the overall project translation, specifically discussing the type-driven phase.
The objective of this phase is to generate a \emph{type-compatible project translation}.

The type-driven translation phase follows Algorithm~\ref{alg:type-driven}.
We start from a Go project $P$ that gets partitioned and the code fragments get organized into a dependency graph by the
\textbf{DependencyAnalysis} function.
We then traverse the resulting graph in post order according to \textbf{PostOrder}.
For each code fragment  \gocode{D}, 
we apply \textbf{FeatureMapping} (as described in Algorithm~\ref{alg:feature-mapping}), \textbf{CompilationCheckAndRepair} (as described in Section~\ref{sec:compilation-repair}) and we check type-compatibility with respect to the execution snapshots extracted by the \textbf{ExecutionSnapshotsCollector} from the project's unit tests (the type-compatibility check is discussed in Section~\ref{sec:type-compatibility} and the execution snapshots collector is described in Section~\ref{sec:implementation}).
If the resulting translation, \rustcode{target\_code}, is both compilable and type-compatible, we store it in the $\mathit{translations}$ map and move to the next code fragment. Otherwise, if we reached the maximum number of tries and the translation is type-compatible, we mock the current function and move to the next fragment (mocking is described in Section~\ref{sec:mocking}). Type-compatibility provides a principled approach for mocking a function and proceeding with the translation of the rest of the project when the LLM encounters difficulties. This approach is only feasible if the generated signature is compatible with that of the source function.
If the current translation is not type-compatible, we have no other choice than to abort the translation. In our experimental evaluation, this situation did not arise (except during the ablation study).

\begin{algorithm}
\caption{Type-driven translation phase}\label{alg:type-driven}
\begin{algorithmic}
  \footnotesize
\Require Source Go Project $P = \overline{\gocode{D}}$, Budget $\mathit{max\_tries}$, Budget $\mathit{requery\_budget}$, $P$'s test suite $\mathit{test\_suite}$
\Ensure  $\mathit{translations}$, Which Maps Code Fragments, $\overline{\gocode{D}}$, to Corresponding Type-Compatible Rust Translations
\State $\textit{translation\_order} \gets \textbf{PostOrder}(\textbf{DependencyAnalysis}(P))$
\State $\textit{translations} \gets \{\}$
\For{$\gocode{D} \in \textit{translation\_order}$}
\State $\mathit{budget} \gets \mathit{max\_tries}$
\While{$\mathit{true}$}
\State $\rustcode{target\_code} \gets \textbf{FeatureMapping}(\gocode{D}, \mathit{requery\_budget})$
\State $\rustcode{target\_code}, \textit{compiled} \gets \textbf{CompilationCheckAndRepair}(\mathit{translations}+[\gocode{D}: \rustcode{target\_code}])$
\State $\mathit{execution\_snapshots} \gets \textbf{ExecutionSnapshotsCollector}(\gocode{D}, \mathit{test\_suite})$
\State $\mathit{type\_compatible} \gets 
\textbf{TypeCompatibilityCheck}(\gocode{D},  \rustcode{target\_code}, \mathit{execution\_snapshots})$
\If{$\mathit{compiled} \wedge \mathit{type\_compatible}$}
\State {$\mathit{translations}[\gocode{D}] \gets \rustcode{target\_code}$}
\State {\bf break}
\EndIf
\State $\mathit{budget} \gets \mathit{budget}-1$
\If{$\mathit{budget}\leq 0$}
\If{$\mathit{type\_compatible}$}
\State $\textit{translations}[\gocode{D}] \gets \textbf{Mock}(\rustcode{target\_code})$
\State {\bf break}
\EndIf
\State \textbf{AbortTranslation()}
\EndIf
\EndWhile
\EndFor
\end{algorithmic}
\end{algorithm}

\subsection{Compilation check and repair} \label{sec:compilation-repair}
The compilation check and repair routine assembles all previously translated fragments into a Rust project that can be sent to the Rust compiler. This task is complex, as real-world Go projects often have their source code spread across intricate project layouts, and Go and Rust follow different organizational principles. In Go, a project is composed of packages that may span multiple files, with private items accessible within the same package even if they are in different files. In contrast, Rust follows a module system where each file in a \textcolor{Rust}{cargo} project is treated as a separate module, and items in one module are not directly accessible in another. Generating appropriate import statements is challenging and has been reported as a leading source of compilation errors in LLM-based translations~\cite{yin2024rectifiercodetranslationcorrector}.

\emph{\textbf{Internal imports.}} 
We symbolically generate the appropriate import statements by using our dependency analysis, and amend the translated code accordingly.

\emph{\textbf{External imports.}} Translation of Go code that uses libraries (either standard or 3rd party) is challenging because we either need to translate the entire library, or we must find appropriate Rust libraries that have the same functionality as their Go counterpart.
Our solution is to rely on the LLM to find appropriate library mappings -- our experience is that, owing to their large training sets, LLMs are frequently able to identify suitable libraries. Then, we symbolically generate the corresponding external import statements, and add them to the translation.

\emph{\textbf{Compilation repair.}} If the compilation check fails, we requery the LLM to obtain a repair for the current fragment being translated. This repair process is localized, restricting changes to only the fragment in question and preventing modifications to other parts of the project. The requerying process leverages error messages from the Rust compiler and follows the approach outlined in~\cite{deligiannis2023fixing}.

\subsection{Type-compatibility} \label{sec:type-compatibility}
The usage of a type within a project depends on local preconditions at its use sites. For instance, in Go, the \gocode{int} primitive type represents a platform-dependent signed integer. To fully preserve semantics, \gocode{int} should map to \rustcode{isize} in Rust. However, \gocode{int} is often used for array indexing in Go, where \rustcode{usize} is typically used in Rust. Depending on the local preconditions at the use sites (i.e. whether \gocode{int} is used for array indexing at all occurrences), the appropriate translation could be either \rustcode{usize} or \rustcode{isize}.

Instead of enforcing strict type equivalence between the original and translated types, we introduce the concept of ``type-compatibility'', inspired by inter-language interoperability~\cite{multi-language-semantics}, which addresses scenarios where values are sent over the boundary between two languages.
The core idea is that any feasible value that can inhabit a type in the original Go project should also be able to inhabit the corresponding type in the translated Rust project. By ``feasible'', we mean values that adhere to the use-site preconditions in the source Go code. These preconditions could, in theory, be statically inferred, but doing so is challenging. Therefore, in this work, we ensure value feasibility by selecting only values from the execution snapshots extracted from the project's test suite.

\paragraph{\textbf{Type-compatibility checks}}

We leverage the data marshaling mechanisms available in both languages, given that Go and Rust provide well-maintained libraries for marshaling complex data types. Each Go type \gocode{$T$} has an associated serialization function, $\mathcal{S}_\gocode{$T$}$, which converts a Go value of type \gocode{$T$} into JSON. Similarly, each Rust type \rustcode{$T_r$} has a corresponding serialization function, $\mathcal{S}_{\rustcode{$T_r$}}$.

For both languages, we require fallible deserialization functions---$\mathcal{D}_{\gocode{$T$}}$ for Go and $\mathcal{D}_{\rustcode{$T_r$}}$ for Rust---which take JSON and return a value of the underlying type (\gocode{$T$} and \rustcode{$T_r$}, respectively). If the JSON object does not conform to the underlying type, the deserialization functions return an error. These serialization and deserialization functions are expected to satisfy the following round-tripping property:
    $(\mathcal{D}_\gocode{$T$} \circ \mathcal{S}_\gocode{$T$})(v) \equiv v$
and $(\mathcal{D}_\rustcode{$T_r$} \circ \mathcal{S}_\rustcode{$T_r$})(v) \equiv v$.
In the other direction, given a JSON object j, if $\mathcal{D}_\gocode{$T$}\left(j\right)$ and $\mathcal{D}_\rustcode{$T_r$}\left(j\right)$ do not abort with an error, we have
    $(\mathcal{S}_\gocode{$T$} \circ \mathcal{D}_\gocode{$T$}) (j) \equiv j$
and
$(\mathcal{S}_\rustcode{$T_r$} \circ \mathcal{D}_\rustcode{$T_r$}) (j) \equiv j$.

\paragraph{Type-compatibility checks for type definitions.}
For each type definition in the translated Rust code, we check type-compatibility with its corresponding Go type definition according to Definition ~\ref{def:type-equiv}. 
\begin{definition}\label{def:type-equiv} \emph{(Type-compatibility)}
    Given a type \gocode{$T$} in the source code, a set \gocode{$V$} of feasible values
    of type \gocode{$T$} and a target type \rustcode{$T_r$}, we say that \rustcode{$T_r$} is compatible with \gocode{$T$} with respect to \gocode{$V$} (written as 
    $\rustcode{$T_r$}\Lleftarrow_{\gocode{$V$}} \gocode{$T$}$)
    if any $\gocode{$v$} \in \gocode{$V$}$ 
    can cross the boundary to the target language as 
    \rustcode{$v_r$} of type \rustcode{$T_r$} such that 
    \rustcode{$v_r$} = $(\mathcal{D}_\rustcode{$T_r$} \circ \mathcal{S}_\gocode{$T$})\left(\gocode{$v$}\right)$ and 
    $\gocode{$v$} = (\mathcal{D}_\gocode{$T$} \circ \mathcal{S}_\rustcode{$T_r$})\left(\rustcode{$v_r$}\right)$.
\end{definition}

\paragraph{Type compatibility checks for function and method signatures.}
For each function declaration in the translated Rust code, we check type-compatibility between its signature and the one of its Go counterpart according to Definition~\ref{def:signature-equiv}.

\begin{definition}[Type-compatible function signatures] \label{def:signature-equiv}
    Given a function \gocode{func $f(\overline{x ~T})~ \overline{U}~ \{ ... \}$} in the source code, sets \gocode{$\overline{V_T}$} of feasible values
    for types \gocode{$\overline{T}$}, sets \gocode{$\overline{V_U}$} of feasible values
    for types \gocode{$\overline{U}$} and a target function \rustcode{fn $f(\overline{x: ~T_r}) ~$->$~ \overline{U_r}~ \{ ... \}$},
    we say that the target function is type-compatible with its source counterpart if the following hold:
    \begin{itemize}
        \item \rustcode{$\overline{T_r}$} is respectively type-compatible to \gocode{$\overline{T}$} with respect to \gocode{$\overline{V_T}$} 
        \item \rustcode{$\overline{U_r}$} is respectively type-compatible to \gocode{$\overline{U}$} with respect to \gocode{$\overline{V_U}$}.
\end{itemize}
\end{definition}

The type-compatibility check for method signatures is similar to the one for functions, with the addition that the types of the receivers must also be compatible. 

For brevity, throughout this paper, we use ``type-compatibility'' to indicate that a Rust entity \rustcode{$B$} is type-compatible with a Go entity \gocode{$A$} based on the execution snapshots. 

\paragraph{\textbf{Type-compatible project translation}}

At the end of the type-driven phase, we expect to obtain a project translation that is type-compatible with the source project based on the feasible values obtained from the execution snapshots, as defined next. 

\begin{definition}[Type-compatible project translation]
Given a source project \gocode{$P$} and its translation \rustcode{$P_r$}, \rustcode{$P_r$} is type-compatible with \gocode{$P$} if:
\begin{itemize}
    \item[i] For each two corresponding type definitions \gocode{$T$} and \rustcode{$T_r$} from \gocode{$P$} and \rustcode{$P_r$}, respectively, \rustcode{$T_r$} is type-compatible with \gocode{$T$} (with respect to the values collected from the unit tests).
\item[ii] For each two corresponding functions/methods \gocode{$f$} and \rustcode{$f_r$} from \gocode{$P$} and \rustcode{$P_r$}, respectively, \rustcode{$f_r$}'s signature is type-compatible with \gocode{$f$}'s signature (with respect to the values collected from the unit tests).
    \item[iii] Project \rustcode{$P_r$} passes the target Rust compiler.
\end{itemize}
    
\end{definition}

\subsection{Function mocking} \label{sec:mocking}
Certain idioms allowed in the source code are strictly disallowed in the target language, making it challenging for the LLM to produce a viable translation. This poses a significant issue when translating large codebases, as encountering such cases can halt the entire translation process.

\begin{figure}[!htb]
    \centering
    \begin{minipage}{.5\textwidth}
        \begin{lstlisting}[language=go, basicstyle=\tiny\ttfamily,escapechar=\&,,backgroundcolor=\color{GoBackground}]
func updateRanks(ranks *Rank, algorithm Algorithm) {
    for _, word := range ranks.Words {
        weight := algorithm.WeightingHits(word.ID, ranks)
        word.Weight = weight
        ...
    }
    ...
}
&\newline&
\end{lstlisting}
        \caption{Source Go snippet}
        \label{fig:go_bad_source}
    \end{minipage}%
    \begin{minipage}{0.5\textwidth}
        \begin{lstlisting}[language=rust, basicstyle=\tiny\ttfamily,]
pub fn update_ranks(ranks: &mut Rank, algorithm: &dyn Algorithm) {
    for word in ranks.words.values_mut() {
        // Failed to compile
        let weight = algorithm.weighing_hits(word.id, ranks);
        word.weight = weight;
        ...
    }
}
\end{lstlisting}
        \caption{Incorrect translation to Rust}
        \label{fig:rust_bad_translation}
    \end{minipage}
\end{figure}

\begin{figure}[!htb]
    \centering
    \begin{minipage}{0.65\textwidth}
        \begin{lstlisting}[language=rust, basicstyle=\tiny\ttfamily]
pub fn update_ranks(ranks: &mut Rank, algorithm: &dyn Algorithm) {
    extern "C" {
        // import original Go function directly
        fn updateRanks(..)
    }
    return updateRanks(..)
}
\end{lstlisting}
        \caption{Function mock}
        \label{fig:rust-function-mock}
    \end{minipage}
\end{figure}
For illustration, the Go snippet in Figure~\ref{fig:go_bad_source}
is particularly challenging to translate to Rust as it violates invariants demanded by the Rust's borrow checker. In particular, the loop condition \gocode{for \_, word := range ranks.Words} conceptually mutably borrows \gocode{ranks.Words} since the loop body performs mutations through \gocode{word}. However, in the loop body, \gocode{ranks} is used for the second time: \gocode{algorithm.WeightingHits(word.ID, ranks)}. This is similar to the  \emph{iterator invalidation} problem in C++, and it is not allowed in Rust. LLMs tend to 
follow the syntax of the original Go snippet and generate the Rust code in Figure~\ref{fig:rust_bad_translation}, which doesn't pass the Rust compiler.

Our observation is that the translation process should be able to bypass such scenarios where the LLM struggles to find a compilable translation, and to continue with the translation of the rest of the project. To address this, we define \emph{function mocks}.
Such a mock consists of a Rust function signature that passes our type-compatibility check, and a function body that calls the original Go function as-is. For the current example, the mock is given in \autoref{fig:rust-function-mock}. 
This necessitates the Go-Rust boundary notions mentioned before, as we need to transform Rust inputs into Go inputs, and the Go output to the Rust output. Assuming $\gocode{func}\ f\ (x\ T)\ U\ \{\dots\}$ is translated into $\rustcode{fn}\ g\ (y: T_r)\rightarrow U_r\ \{\dots\}$, we define the function mock for $g$ to be:
$\textbf{Mock}(g) ::= \mathcal{D}_{U_r}\circ\mathcal{S}_U\circ f\circ\mathcal{D}_T\circ\mathcal{S}_{T_r}$. Notably, function mocking is only possible if we manage to obtain a function signature that passes our type-compatibility checks. 

\section{Semantics-Driven Translation}\label{sec:semantics}
 The objective of this stage is to refine the type-compatible project translation produced by the type-driven translation so that it becomes I/O equivalent to the original. 
The semantics-driven translation phase follows Algorithm~\ref{alg:semantics-driven}.
For each translation in the $translations$ map produced by the type-driven translation, \textbf{I/OEquivalenceCheck} checks 
whether it is I/O equivalent with the original 
Go fragment on the execution snapshots extracted by the \textbf{ExecutionSnapshotsCollector} from $P$'s unit tests.
Notably, this check is local, only focusing on the semantic correctness of the current fragment. To achieve this, we construct mocks for all its callees (if applicable), which emulate the behavior of the original code as described in Section~\ref{sec:mocking}.
If the check succeeds, then we exit the \textbf{while} loop and update the $translations$ map.
Otherwise, we attempt to re-generate and refine the body of the current function by calling \textbf{FeatureMapping} and 
\textbf{CompilationCheckAndRepair}, both
with $\mathit{freezing\_signature}$ set to $\mathit{true}$ so that they are disallowed from changing types. We keep refining and checking I/O equivalence until we obtain a semantically correct translation or we run out of budget. On exit from the \textbf{while} loop, the translation is guaranteed to be type-compatible (as the LLM was disallowed from modifying the signature), but it may still not be I/O equivalent to the original fragment. 

\subsection{I/O equivalence check}

We define two types of semantic checks for a Go project $P$: one for global variable initializations and another for functions. Our I/O equivalence check operates under the assumption that the target Rust code adheres to the structure specified by the feature mapping in Section~\ref{sec:feature-mapping}. As we need to pass values between Go and Rust, we make use of the serialization and deserialization functions introduced in Section~\ref{sec:type-compatibility}. 

\begin{definition}[I/O equivalence of global variable initialization]\label{def:semantic-global-var}
If we have $\gocode{D}: \gocode{var}\ x\ T = f(\overline{v})$, that gets translated to  $\textit{translations}[\gocode{D}]$, 
we say that \gocode{D} and $\textit{translations}[\gocode{D}]$ 
are I/O equivalent if
$\textit{translations}[\gocode{D}]$ is of the form $\rustcode{static}\ y: \rustcode{Lazy<}U\rustcode{>} = \rustcode{Lazy::new(||} \dots \rustcode{)}$ and   
$\mathcal{S}_T(x) = \mathcal{S}_U(\rustcode{Lazy::force(}y\rustcode{)})$.
\end{definition}

\rustcode{Lazy::force} forces the evaluation of the static var $y$.
The left and right hand sides of the equality check in the definition are serialized to JSON objects, which allows for plain string comparison.
\paragraph{I/O equivalence of functions}
Let's consider the Go function $\gocode{D}: \gocode{func}\ f\ (\overline{x\ T})\ (\overline{U}, \gocode{error}) \in P$, which has inputs $\overline{i}$, outputs $\overline{o}$, and
an additional error output $\mathit{err}$.  
By the (Map-Error-Handling-Fn) feature mapping rule in Section~\ref{sec:error-handling}, we expect $\textit{translations}[\gocode{D}]$ to be of the form $\rustcode{fn}\ g\ (\overline{y:Tr}) \rightarrow \rustcode{Result<}\overline{Ur}, \rustcode{anyhow::Error>}$. 
We check I/O equivalence between functions/methods with respect to a set of values $V=\{(\overline{i}, \overline{o'}, err)^*\}$ collected by running the unit tests in the original Go repository.
We overload the output by considering $\overline{o'}$ to be an extension of the actual output $\overline{o}$ that accounts for possible side-effects.

\begin{definition}[I/O equivalence of functions]
If we have $\gocode{D}: \gocode{func}\ f\ (\overline{x\ T})\ (\overline{U}, \gocode{error})$, we say that \gocode{D} and $translations[\gocode{D}]$ are I/O equivalent with respect to $V$ if $translations[\gocode{D}]$ is of the form 
$\rustcode{fn}\ g\ (\overline{y: Tr}) \rightarrow \rustcode{Result<}\overline{Ur}, \rustcode{anyhow::Error>}$ and
\begin{eqnarray*}
\bigwedge
    \begin{array}{l}

        err \equiv \gocode{nil} 
        \rightarrow 
        \rustcode{Ok}(\overline{res}) = g(\mathcal{D}_{\overline{Tr}}(\mathcal{S}_{\overline{T}}(\overline{i}))) \wedge
        \mathcal{S}_{\overline{Ur}}(\overline{res}) \equiv \mathcal{S}_{\overline{U}}(\overline{o'})
         \\
        err \not\equiv \gocode{nil} \rightarrow  \rustcode{Err}(\_) = g(\mathcal{D}_{\overline{Tr}}(\mathcal{S}_{\overline{T}}(\overline{i})))
    \end{array}
\end{eqnarray*}
\end{definition}

We abuse the notation using $S_{\overline{T}}(\overline{i})$ to mean 
pairwise application of the corresponding serialization function to each individual input in the tuple $\overline{i}$.
The definition says that, for any input $\overline{i}$, 
if the Go function $f$ succeeds on $\overline{i}$, then so should the Rust translation $g$ on the corresponding input $\mathcal{D}_{\overline{Tr}}(\mathcal{S}_{\overline{T}}(\overline{i}))$, and the output states should be the same, 
$\mathcal{S}_{\overline{Ur}}(\overline{res}) \equiv \mathcal{S}_{\overline{U}}(\overline{o'})$. Alternatively, if $f$ fails, then so should $g$.

\begin{algorithm}
\caption{Semantics-driven translation phase}\label{alg:semantics-driven}
\begin{algorithmic}
  \footnotesize
\Require $\textit{translations}$ Maps Code Fragments, $\overline{\gocode{D}}$, to Type-Compatible Rust Translations, Budget $\mathit{requery\_budget}$, Budget $max\_tries$, $P$'s test suite $\mathit{test\_suite}$
\Ensure  $\textit{translations}$ Maps code fragments, $\overline{\gocode{D}}$, to Their Final Rust Translations
\For{$\gocode{D}, \rustcode{target\_code} \in \textit{translations}$}
\State $\mathit{budget} \gets \mathit{max\_tries}$
\While{$\mathit{budget}>0$}
\State $\mathit{execution\_snapshots} \gets \textbf{ExecutionSnapshotsCollector}(\gocode{D}, \mathit{test\_suite})$
\State $\textit{equivalent} \gets \textbf{I/OEquivalenceCheck}(\gocode{D}, \rustcode{target\_code}, \mathit{execution\_snapshots})$
\If{$\textit{equivalent}$}
\State {\bf break}
\EndIf
\State $\mathit{compiled} \gets \mathit{false}$
\While{$\neg \mathit{compiled} \wedge \mathit{budget} > 0$}
\State $\rustcode{target\_code} \gets \textbf{FeatureMapping}(\gocode{D}, requery\_budget, \textit{freezing\_signature}=\mathit{true})$
\State $\rustcode{target\_code}, \textit{compiled} \gets \textbf{CompilationCheckAndRepair}(\rustcode{target\_code}, \textit{freezing\_signature}=\mathit{true})$
\State $\mathit{budget} \gets \mathit{budget} - 1$
\EndWhile
\EndWhile
\State $\textit{translations}[\gocode{D}] \gets \rustcode{target\_code}$
\EndFor
\end{algorithmic}
\end{algorithm}

\section{Experimental Evaluation}
\label{sec:evaluation}
We conduct an evaluation of \ourtool{} to assess the following research question:

\newcommand{\rqone}[0]{How effective is \ourtool{} at translating entire projects?}
\newcommand{\rqtwo}[0]{How much do our proposed type-compatibility checks and feature mapping rules benefit translation?}
\newcommand{\rqthree}[0]{How does \ourtool{} compare to parallel works in whole-repository translation?}
\begin{itemize}
    \item \rqone{} Specifically, how much of the translated code can compile, and how many of the functions can be validated I/O equivalent?
    \item \rqtwo{}
    \item \rqthree{}
\end{itemize}

\subsection{Experimental Setup}
\subsubsection{Implementation} \label{sec:implementation}
\ourtool{} takes as input (1) a \go{} project, (2) $\mathit{requery\_budget}$ for \autoref{alg:feature-mapping}, (3) $\mathit{max\_tries}$ for \autoref{alg:type-driven}, and (4) $\mathit{max\_tries}$ for \autoref{alg:semantics-driven}.
\ourtool{} can be configured to use both proprietary and open-source LLMs. \ourtool{} outputs a \rust{} translation of the input project. The translation may have some function/method bodies replaced with mocks as described in Section~\ref{sec:type-driven}. The core code modules of \ourtool{} are (1) the execution snapshot collector, (2) the project partitioner, (3) the type-driven translator, and (4) the semantics-driven translator. All modules are implemented in python, and use py-tree-sitter for parsing and instrumenting code. The project partitioner, type-driven translator, and the semantics-driven translator implement the algorithms described in the previous sections, so we focus on the execution snapshot collector, and testing I/O equivalence.

The execution snapshot collector collects inputs and expected outputs for functions in the source project. It first instruments each function in the source project with statements that log the inputs and outputs of all functions in JSON format. It then executes the unit tests of the source project to collect these inputs and outputs. These input-output examples are used to test type-compatibility during type-driven translation, and they are used to create unit tests for the I/O equivalence check in the semantics-driven translation.

Before performing semantics-driven translation, \ourtool{} creates a Rust unit test for each function in the Rust translation. For a given function in Rust, its unit test first loads all input-output examples collected for the corresponding function in the source project using JSON deserialization. It then executes all loaded inputs on the Rust function, and compares the computed outputs to the expected outputs that were loaded. A unit test passes if and only if all computed outputs match the expected outputs. If we are not able to collect any input-output examples for a given function, we create an empty unit test that automatically fails. The percentage of all unit tests that pass is equivalent to the percentage of functions that are validated I/O equivalent.

\subsubsection{LLMs}
We use Anthropic's Claude 3 Sonnet~\cite{claude} provided by Amazon Bedrock. Prior work~\cite{eniser2024translatingrealworldcodellms} has shown that Claude 3 Sonnet performs similarly to other state-of-the-art proprietary LLMs, such as GPT-4o~\cite{achiam2023gpt} and Gemini Pro~\cite{gemini}, thus we believe our results apply to them as well. In order to enable others to reproduce our results deterministically without the need to pay to run the LLM, \ourtool{} logs the inputs and outputs of the LLM, and supports replaying these logs.

\subsubsection{Benchmarks}
We use real-world projects collected from GitHub as our benchmarks. Our main criteria for selecting projects are that (1) the project has more than 100 stars on GitHub or is actively maintained (specifically, the project has commits in the last 6 months) and (2) the project only makes use of \go{} standard libraries. As explained in Section~\ref{sec:compilation-repair}, our approach supports generating code that utilizes third-party libraries and we already do so for standard Go libraries (e.g. regex) that don't always have Rust standard library correspondents and instead require Rust 3rd party libraries. However, for this evaluation, we selected Go benchmarks that do not rely on such libraries as standard Go libraries are more likely to have equivalent libraries (including third-party options) available in Rust. The benchmarks we select are listed below.

\begin{itemize}
    \item \textbf{ach}~\cite{ach}: a Go library implementing a reader, writer, and validator for banking operations. We translate a sub-module of this project that implements validation logic.
    \item \textbf{go-edlib}~\cite{gohyphenedlib}: a Go library string comparison and edit distance algorithms
    \item \textbf{stats}~\cite{stats}: a Go library implementing statistical algorithms
    \item \textbf{textrank}~\cite{textrank}: a TextRank implementation in Golang with extendable features (summarization, phrase extraction) and multi-threading
    \item \textbf{histogram}~\cite{histogram}: a Go library implementing streaming approximate histograms
    \item \textbf{gonameparts}~\cite{gonameparts}: provides string algorithms for parsing human names into parts
    \item \textbf{checkdigit}~\cite{checkdigit}: provides check digit algorithms and calculators
\end{itemize}

\begin{table}[]
\caption{Benchmark details}
\label{tbl:benchmarkdetails}
\centering
\resizebox{0.8\textwidth}{!}{
\begin{tabular}{@{}rccccccc@{}}
\toprule
Project & LoC & \makecell{\# Functions/\\Methods} & \makecell{\# Structs/\\Interfaces} & \makecell{Unit Test\\Statement Coverage} & Stars & Forks \\ \midrule
\textit{ach}        & 6642 & 369 & 75 & 92.9\% & 442 & 145 \\
\textit{go-edlib}   &  639 & 25  & 1  & 100\% & 480 & 24 \\
\textit{stats}      & 1241 & 79  & 8  & 93.2\% & 2.9K & 168 \\
\textit{textrank}   & 1132 & 69  & 20 & 94.6\% & 205 & 22 \\
\textit{histogram}  &  314 & 23  & 5  & 43.2\%  & 175 & 31 \\
\textit{gonameparts}&  413 & 15  & 2  & 96.1\%  & 42  & 5 \\
\textit{checkdigit} &  428 & 29  & 9  & 100\%  & 110 & 7 \\
\bottomrule
\end{tabular}
}
\end{table}

Details about our benchmarks are given in Table~\ref{tbl:benchmarkdetails}. For each benchmark, we report the total lines of code (excluding unit tests), the number of functions and methods defined in the project, the number of structs defined, the statement coverage achieved by the project's unit tests, and the number of stars and forks on GitHub.

\subsubsection{\ourtool{} Parameters}
In general, we set $\mathit{requery\_budget}$ and both $\mathit{max\_tries}$ parameters high enough to the point that we reach diminishing returns (i.e., setting them higher would rarely yield better results). Specifically, we set $\mathit{requery\_budget}$ to 10, $\mathit{max\_tries}$ for \autoref{alg:type-driven} to 15, and $\mathit{max\_tries}$ for \autoref{alg:semantics-driven} to 5.
We use a relatively low temperature (i.e., less random) of 0.2 for Claude 3 Sonnet to produce more reliable and deterministic results.

\subsection{Results}

\begin{table}[]
\caption{Translation results for \ourtool{}}
\label{tbl:results}
    \centering
    \resizebox{\textwidth}{!}{
    \begin{tabular}{c|cc|cc|cc}
        \hline
        \multirow{2}{*}{\textbf{Benchmark}} & \multicolumn{2}{c|}{\textbf{Full}} & \multicolumn{2}{c|}{\textbf{No Type Check}} & \multicolumn{2}{c}{\textbf{No Feature Map}} \\
        \cline{2-7}
        & \textbf{\% Compiled} & \textbf{\% Equivalent} & \textbf{\% Compiled} & \textbf{\% Equivalent} & \textbf{\% Compiled} & \textbf{\% Equivalent} \\
        \hline
\textit{ach}            & 96    & 65  & 93  & 62 & 5  & 0 \\
\textit{textrank}       & 97    & 75  & 97  & 67 & 16 & 0 \\
\textit{go-edlib}       & 100   & 81  & 100 & 81 & 20 & 0 \\
\textit{stats}          & 100   & 73  & 99  & 54 & 3 & 0 \\
\textit{histogram}      & 100   & 63  & 100 & 63 & 96 & 0 \\
\textit{gonameparts}    & 100   & 71  & 100 & 29 & 29 & 0 \\
\textit{checkdigit}     & 100   & 86  & 100 & 76 & 21 & 0 \\
        \hline
        \hline
Average                    & 99    & 73  & 98  & 61  & 27 & 0
    \end{tabular}
    }
\end{table}

\paragraph{\rqone}
We run \ourtool{} on each of our benchmarks, and report results in Table~\ref{tbl:results} under the column \textbf{Full}. We report two key metrics: compilation success rate under the column \textbf{\%~Compiled} and function equivalence rate under the column \textbf{\%~Equivalent}. Compilation success rate reports the amount of the source project's code we are able to translate into compiling Rust code, measured as $100\times\frac{\textit{compiling\;lines\;of\;code}}{\textit{total\;lines\;of\;code}}$. Function equivalence rate measures the percentage of functions that are I/O equivalent on all collected input-output examples. Importantly, this rate depends on the coverage of the test suite: functions not covered by unit tests are automatically marked as not equivalent. For example, in \textit{histogram}, one-third of the functions did not have unit test coverage, preventing us from validating their equivalence to the translated versions.

Our results show that \ourtool{} produces code that compiles nearly 100\% of the time, which is a substantial challenge itself given the restrictiveness of \rust{}'s type system. In addition, \ourtool{} raises the bar in producing semantically correct code. On average 73\% of functions are I/O equivalent to the source project. This shows that \ourtool{} can significantly reduce developer effort to translate a project. Prior work~\cite{ibrahimzada2024repositorylevelcompositionalcodetranslation} has shown that even a translation where only $\sim$45\% of functions are I/O equivalent significantly reduces the amount of developer effort to translate a project.

\paragraph{\rqtwo}
We run \ourtool{} with type-compatibility checks disabled (but with feature mapping enabled), and with both type-compatibility checks and feature mapping disabled. The compilation success rates and function equivalence rates are reported under the columns \textbf{No Type Check} and \textbf{No Feature Mapping}, respectively, in \autoref{tbl:results}. 

Our results shows that our type-compatibility checks are generally helpful. They improve function equivalence rate for five out of our seven benchmarks, usually by 5--35\%, and in one case by 144\%. In addition, our feature mapping rules are absolutely critical for reliability. Without them, \autoref{alg:type-driven} aborts for all of our benchmarks, preventing us from reliably evaluating function equivalence. The sub-column \textbf{\% Compiled} under the super-column \textbf{No Feature Mapping} reports the percentage of functions that were successfully translated before \autoref{alg:type-driven} aborted.

\paragraph{\rqthree}
To the best of our knowledge, only two works~\cite{ibrahimzada2024repositorylevelcompositionalcodetranslation, shiraishi2024contextawarecodesegmentationctorust}, done in parallel to ours, tackle whole-project  translation. Neither of these tools support Go as source language, and only one~\cite{shiraishi2024contextawarecodesegmentationctorust} supports Rust as a target language, thus directly running their tools on our benchmarks would be impractical. However, certain aspects of the results reported in these works can be directly compared to ours, and we believe our work compares favorably to them.

Both sets of authors report very high success rates (100\% and 99\%) of getting ``runnable'' (i.e. syntactically valid and/or compiling) translations, which is similar to \ourtool{} at 99\%. However, \ourtool{} is much more successful at producing validated I/O-equivalent translations. The first work~\cite{shiraishi2024contextawarecodesegmentationctorust} is unable to validate semantic correctness for the vast majority of the translated code because the test cases crash. This is mostly because they do not attempt to robustly handle semantic correctness. The second work~\cite{ibrahimzada2024repositorylevelcompositionalcodetranslation} does attempt to robustly handle I/O equivalence. They report that they can validate I/O equivalence of 26\%
of the functions they translate on average (corresponding to 46\% of the functions actually covered by unit tests), which is substantially less than our 73\%. Moreover, they report that semantic validation simply crashes 25\% of the time. We believe the difference in performance can be credited to our type-compatibility checks, which ensure that semantic validation does not crash, allowing us to successfully validate many more functions.

\subsection{Discussion}

$\indent$\emph{Remaining non-compilable code.} 
\ourtool{} predominantly generates compilable translations. For the remaining non-compilable functions, we found that the LLM often struggles to generate code that requires deviations from the original syntax. However, this is sometimes necessary in order to satisfy Rust's strict compiler requirements. For instance, the LLM frequently translates use-after-move patterns, such as \gocode{f(x); y = x.field.Clone()}, directly into Rust equivalents, leading to compilation errors. A compilable translation would require modifying the code to clone \gocode{x.field} first and then call \gocode{f}, deviating from strict syntax similarity. We observed that these syntax-altering patterns accounted for 53\% of the non-compilable functions in \textit{ach}.

\emph{Remaining inequivalent code.} 
\ourtool{} achieves a high I/O equivalence rate. For the remaining inequivalent functions, our manual investigation identified semantic errors such as calling incorrect functions or using inappropriate operators. Additionally, some functions exhibit nondeterministic behavior, making equivalence validation particularly challenging. For example, in \textit{stats}, there is a code snippet \gocode{perm := rand.Perm(length)} that generates a random permutation of the slice $[0..length)$. While \ourtool{} correctly translates this code, we were unable to validate equivalence to the original code due to the inherent randomness.

\section{Related Work}
\label{section:related_work}
In this section, we discuss closely related work from the literature under several categories. 

\textbf{Entire Project Translation.}
The most closely related work to ours are two parallel works on translating entire projects using LLMs~\cite{ibrahimzada2024repositorylevelcompositionalcodetranslation, shiraishi2024contextawarecodesegmentationctorust}. The first work~\cite{ibrahimzada2024repositorylevelcompositionalcodetranslation} targets translating Java to Python. They propose a partitioning technique and I/O equivalence validation technique that is similar to ours. They also propose to build a symbolic rule-based mapping for APIs in the source language to APIs in the target language, which is somewhat similar to our feature mapping, though we argue it is less flexible than ours. In addition, their technique for I/O equivalence validation is less reliable than ours, as they often report failures to map concrete values in the source language to the target language. The second work~\cite{shiraishi2024contextawarecodesegmentationctorust} does not propose technique for semantic validation at all. Their translation approach resembles ours, but without feature mapping or type compatibility checks. The only other works known to us on entire project translation are based purely on symbolic rules. They target C to Rust,~\cite{zhang2023ownership, emre2021translating,c2rust}, C to Go~\cite{cgo}, and Java to C\#~\cite{sharpen}. However, purely symbolic rule based translation is known for producing unidiomatic code~\cite{eniser2024translatingrealworldcodellms}.

\textbf{Other Code Translation Works.}
The vast majority of other work on code translation~\cite{tang-etal-2023-explain, jiao2023evaluation, RoziereLachaux2020, RoziereZhang2022, szafraniec2022code,jana2023attention,yang2024vertverifiedequivalentrust,codefuse,structcoder,yan2023codetransocean, yin2024rectifiercodetranslationcorrector} has focused on translating code taken from competitive programming websites~\cite{puri2021codenet, codexglue}, educational websites~\cite{ahmad2021avatar, yan2023codetransocean}, or hand-crafted coding problems~\cite{liu2024your,chen2021evaluating}. There are a few exceptions~\cite{PanICSE24,eniser2024translatingrealworldcodellms,zhang2023multilingual}, but they report very little success on translating code exceeding 100 lines of code.
Many of these works~\cite{RoziereLachaux2020, RoziereZhang2022, szafraniec2022code,jana2023attention,structcoder} propose novel training methodologies for LLMs, which could be applied to the underlying LLM used by our approach to potentially improve performance. Others propose prompting techniques~\cite{tang-etal-2023-explain} and repair techniques~\cite{yin2024rectifiercodetranslationcorrector}, which could be applied to our approach as well.
Also relevant to our are work are those on automated program repair, which can be leveraged to repair I/O equivalence errors. Several recent works~\cite{xia2023automated,kong2024contrastrepair} use LLMs to propose repairs, though their techniques would likely need to be adapted to repair translation errors.

\textbf{Cross Language Differential Testing and Verification.}
While prior work has proposed techniques for cross-language differential testing~\cite{eniser2024translatingrealworldcodellms, ibrahimzada2024repositorylevelcompositionalcodetranslation} and verification~\cite{garzella2020xlverify,yang2024vertverifiedequivalentrust}, they do not address the critical challenge ensuring compatibility between the implementations they compare. We aim to do this with our type-compatibility checks. We drew inspiration from work in language interoperability~\cite{semantics-soundness-for-interop}, which targets the case when both languages are compiled to a shared intermediate or target language. Also closely related to code translation is translation validation~\cite{DBLP:conf/tacas/PnueliSS98, necula2000translation}, which is most often used to validate compiler correctness.
There are also many other works that focus on same-language differential testing and verification, such as those that use symbolic execution~\cite{noller2020hydiff, bohme2013regression, palikareva2016shadow, person2011directed} and fuzzing~\cite{guo2018dlfuzz, jin2010automated, nilizadeh2019diffuzz, petsios2017nezha}. We could potentially adapt these techniques to obtain stronger guarantees on I/O equivalence, but they are not directly applicable.



\section{Conclusion}

In this work, we present an approach for translating entire projects, and we demonstrate its application to the translation of Go projects to Rust. We propose two novel strategies, namely feature mapping rules and type compatibility checks, and show empirically on Go projects from Github that they improve the reliability of obtaining a compiling and I/O equivalent translation significantly. This work is the first to deliver translations of entire projects that not only compile but also pass a meaningful share of the projects' test suites.

\bibliographystyle{ieeetr}
\bibliography{reference}

\end{document}